\documentclass[pdflatex,sn-mathphys-num]{sn-jnl}

\usepackage{graphicx}%
\usepackage{multirow}%
\usepackage{amsmath,amssymb,amsfonts}%
\usepackage{amsthm}%
\usepackage{mathrsfs}%
\usepackage[title]{appendix}%
\usepackage{xcolor}%
\usepackage{textcomp}%
\usepackage{manyfoot}%
\usepackage{booktabs}%
\usepackage{algorithm}%
\usepackage{algorithmicx}%
\usepackage{dirtytalk}
\usepackage{algpseudocode}%
\usepackage{listings}%

\theoremstyle{thmstyleone}%
\theoremstyle{thmstyletwo}%

\theoremstyle{thmstylethree}%

\begin{document}

\title[Multichrome Contagions overcome vaccine hesitancy]{The Niche Connectivity Paradox: Multichrome Contagions Overcome Vaccine Hesitancy more effectively than Monochromacy}

\author*[1,2]{\fnm{Ho-Chun Herbert} \sur{Chang}}\email{herbert.chang@dartmouth.edu}

\author[2,3]{\fnm{Feng} \sur{Fu}}\email{feng.fu@dartmouth.edu}
\equalcont{These authors contributed equally to this work.}

\affil*[1]{\orgdiv{Quantitative Social Science}, \orgname{Dartmouth College}, \orgaddress{\postcode{03755}, \state{NH}, \country{USA}}}

\affil[2]{\orgdiv{Department of Mathematics}, \orgname{Dartmouth College}, \orgaddress{ \postcode{03755}, \state{NH}, \country{USA}}}

\affil[3]{\orgdiv{Department of Biomedical Data Science}, \orgname{Geisel School of Medicine at Dartmouth}, \orgaddress{\city{Lebanon}, \postcode{03756}, \state{NH}, \country{USA}}}


\abstract{
The rise of vaccine hesitancy has caused a resurgence of vaccine-preventable diseases such as measles and pertussis, alongside widespread skepticism and refusals of COVID-19 vaccinations during the COVID-19 pandemic. While categorizing individuals as either supportive of or opposed to vaccines provides a convenient dichotomy of vaccine attitudes and behaviors, the issue of vaccine hesitancy is far more complex and dynamic. It often involves wavering individuals (switchers) whose attitudes fluctuate, reflecting shifts in their sentiment toward vaccination and related topics---those who may exhibit pro-vaccine attitudes at one time and anti-vaccine attitudes at another, or vice versa, alongside potential changes in their perspectives on other issues. Here, we identify and analyze multichrome contagions as potential targets for intervention by leveraging a dataset of known pro-vax and anti-vax Twitter users ($n =135$ million) and a large COVID-19 Twitter dataset ($n = 3.5$ billion; including close analysis of $1,563,472$ unique individuals). We identify and reconstruct an evolving multiplex sentiment landscape using top co-spreading issues, characterizing them as monochrome and multichrome contagions, based on their conceptual overlap with vaccination. Against this backdrop, we demonstrate switchers as deliberative rather than simplistic: they are more moderate, engage with a wider range of topics, and occupy more central positions in their networks. Further examination of their information consumption shows that their discourse often engages with progressive issues such as climate change, which can serve as avenues for multichrome contagion interventions to promote pro-vaccine attitudes. Using data-driven intervention simulations, we demonstrate a paradox of niche connectivity, where multichrome contagions with fragmented, non-overlapping communities generate the highest levels of diffusion for pro-vaccine attitudes. Our work offers insights into harnessing synergistic hitchhiking effect of multichrome contagions to drive desired attitude and behavior changes in network-based interventions, particularly for overcoming vaccine hesitancy.
}




\maketitle

\section{Introduction}\label{sec:intro}

Vaccine hesitancy has emerged as a critical public health challenge, particularly during the COVID-19 pandemic, as global vaccination efforts have faced resistance from various communities. Addressing vaccine hesitancy effectively requires analysis at the individual level, and how individuals interact across shared domains, such as online social platforms. Leveraging two datasets with more than 3.6 billion Tweets combined, here we explore the characteristics and communities of vaccine hesitators and ``switchers"---those who oscillate between acceptance and denial--- and how we can devise novel forms of network-based interventions. 

Vaccine hesitancy is influenced by a range of factors, such as scientific knowledge and socio-cultural background~\cite{macdonald2015vaccine}. Skeptical parents often resist the health messages that vaccines save lives, due to the lack of observable counterfactuals~\cite{bloom2014addressing}, while also observing imperfect vaccines~\cite{chen2019imperfect}. Theoretical results also point to trade-offs between vaccine efficacy and disease severity such as the reproductive ratio $R_0$~\cite{hu2024evolutionary}. Especially during the COVID-19 pandemic, trust in healthcare systems, professionals, policymakers, and information sources generated the greatest sources of vaccine hesitancy~\cite{larson2018state}, as demonstrated through cross-country comparisons on institutional trust~\cite{lazarus2022revisiting}. Perceptions of expedited vaccine development and trials, side effects, and perceived low risk of disease also produced resistance~\cite{razai2021covid,brewer2017meta}. These factors diminish vaccine uptake. 

Due to the distrust in institutions, vaccine hesitancy during the COVID-19 pandemic was often framed within the context of political partisanship~\cite{chen2021covid}. However, this binary framework overlooks the multifaceted nature of vaccine attitudes, which are often influenced by sociocultural and ideological factors. In fact, historically, vaccine hesitancy has not been confined to any single ideological spectrum. Before 2015, the archetype of vaccine hesitators often included liberals who espoused naturalistic and anti-establishment beliefs~\cite{lewandowsky2013role}. In Japan, where left-wing ideology predominates~\cite{toriumi2024anti}, vaccine hesitancy has been linked to spirituality, naturalism, and alternative health practices. Latent variables also include moral values~\cite{amin2017association}, which are known to intersect with partisanship. The multiplex interaction of factors suggests that a simplistic political dichotomy fails to capture factors related to vaccine hesitancy, and therefore effective downstream intervention. 

To avoid reliance on institutional messaging, the results of the survey and panel results show social norms are in general more effective than government mandates~\cite{schmelz2021overcoming}, though they still exhibit dynamical interactions~\cite{schmelz2022opposition}. Specifically, peer influence refers to the impact that individuals within a social group or network have on each other's attitudes, behaviors, or decisions~\cite{kandel1992peer}. Peer influence can be described by some theoretical social frameworks. For instance, social learning theory posits that individuals learn behaviors by observing and imitating others~\cite{bandura1977social}, or simply conformity~\cite{hirschi2015social}. Regardless of motive, peer influence has been shown to have a substantial effect on vaccine acceptance, where people are more likely to get vaccinated when they know others are doing so~\cite{moehring2023providing}. Presenting vaccine acceptance as a social norm can be an effective strategy to encourage vaccination, especially among those who are hesitant~\cite{moehring2023providing}. Similarly, underestimation of vaccine acceptance by others may contribute to vaccine hesitancy~\cite{moehring2023providing}.

Indispensable to peer influence are social ties. In comparison to message-based interventions, which are broadcasted from institutional sources and represented as simple star networks, network-based approaches have shown promise in addressing vaccine hesitancy. Prior research on networked interventions that mention access to vaccines, immunization mandates, and patient education have demonstrated potential in increasing vaccine uptake among hesitant populations~\cite{jarrett2015strategies}, despite similar messaging from institutional sources. Peer influence in conjunction with social ties plays a crucial role in vaccine acceptance, which necessitates a social network lens in understanding vaccination decisions~\cite{brunson2013impact}.

Social media platforms are therefore often best suited for health-oriented, network-based interventions. For instance, mask wearing interventions during COVID-19 increased by 111 times when mediated by K-pop celebrities~\cite{chang2023parasocial}. However, social media can also serve as a double-edged sword, especially since COVID-19 vaccine hesitancy further differentiates itself due to its global footprint~\cite{sallam2021covid, betsch2012opportunities}. Moreover, vaccine hesitancy is also found to be highly correlated with those who engage with conspiracy theories~\cite{rathje2022social} and thrive in similar environments~\cite{pertwee2022epidemic}. These posts often mix reliable information with inaccurate or biased messages~\cite{puri2020role}, including suspicion of economic or political motivations, trust in government and healthcare systems~\cite{lazarus2021global}, and opinions of authority figures~\cite{loomba2021measuring}. Misinformation and conspiracy theories when presented in pseudo-scientific language significantly impact COVID-19 vaccine acceptance~\cite{roozenbeek2020susceptibility}, due to peer effects and echo chambers that reinforce pre-existing beliefs~\cite{vosoughi2018spread}. Crucially, there is evidence using Facebook that shows vaccine skeptical content diffuses more effectively and yields higher engagements than pro-vaccine materials~\cite{allen2024quantifying}. 

In this paper, we collected, curated and utilized two extensive Twitter datasets. The first dataset (Dataset 1) is the largest public Twitter dataset on the COVID-19 pandemic~\cite{chen2020tracking}, spanning 30 months from January 28, 2020. With 3.5 billion tweets, the data provides a robust foundation for identifying shifts in user sentiment and understanding how communities' attitudes towards vaccination have evolved throughout the pandemic. The second dataset (Dataset 2) is derived from a large, public dataset on anti-vaxers and pro-vaxers~\cite{muric2021covid} and includes data collected retroactively until October 18, 2020. The dataset includes 250 million tweets dating back to 2009, from 107,000 users, allowing us to identify 5,326,468 inter-user connections. This comprehensive dataset captures known anti-vax/pro-vax behavior over time, enabling us to analyze the evolution of vaccine sentiments (pro, anti, and switchers). These datasets were then augmented to include political ideology and removal of social bots (see Methods). While there are many other social media platforms, our goal was to understand both network structure and attributes relating to users (derived from their full timelines), Twitter emerged as the most suitable platform to study users beyond content~\cite{chang2023liberals}.

Our work has three broad contributions. 
First, we identify key communities adjacent to pro-vaccination attitudes. This expands on filling the aforementioned gap by demonstrating, concretely, areas that could benefit from peer influence-based interventions~\cite{allen2024quantifying}.  Second, we articulate vaccination campaigning strategies that lead to effective network-based interventions by conducting simulations on ecologically-valid, data-driven communities. Using large-scale Monte Carlo simulations, we compare a range of network diffusion scenarios with their expected outcomes beyond traditional, theoretical simulations as a methodological contribution~\cite{chang2018co,chang2019co,fugenschuh2023overcoming}. We find that identifying the adjacent niches/audiences these switchers occupy can serve as ways to prime unconverted communities.
Data-driven simulations further show a paradox of ``niche connectivity,'' wherein fragmented, issue-specific communities can generate the greatest overall diffusion of pro-vaccine sentiments. Third, we characterize the ``hesitators'' or ``switchers'' in terms of their information diet and network centrality. Existing studies often assume static positions of individuals as either pro-vaccine or anti-vaccine, and our goal is to fill the gap in terms of dynamic, oscillatory vaccine attitudes. These individuals mirror the oscillatory behaviors observed in other domains, such as independent voters in political systems, who swing between parties and have been characterized as both unsophisticated (low political knowledge) and sophisticated (deliberate decision-makers)~\cite{campbell1980american,converse1964belief}. Vaccine hesitancy could benefit from such a view. These findings highlight the power of cross-issue connectivity in shaping vaccine attitudes and offer targeted strategies for network-based interventions to overcome vaccine hesitancy.

\section{Results}\label{sec:results}

\subsection{Identifying network communities for potential intervention}

In determining the right communities for the intervention, we must first identify the modes of targeted messaging. This includes identifying a) communities and b) individuals for downstream conversion. Prior research has found that identifying individuals with the most followers to be the most effective, and we include a proof in Appendix~\ref{secA3}. We first focus our attention on identifying the correct communities, then consider individuals beyond simple metrics of degree centrality.

\begin{figure}
    \centering
    \includegraphics[width=1.0\linewidth]{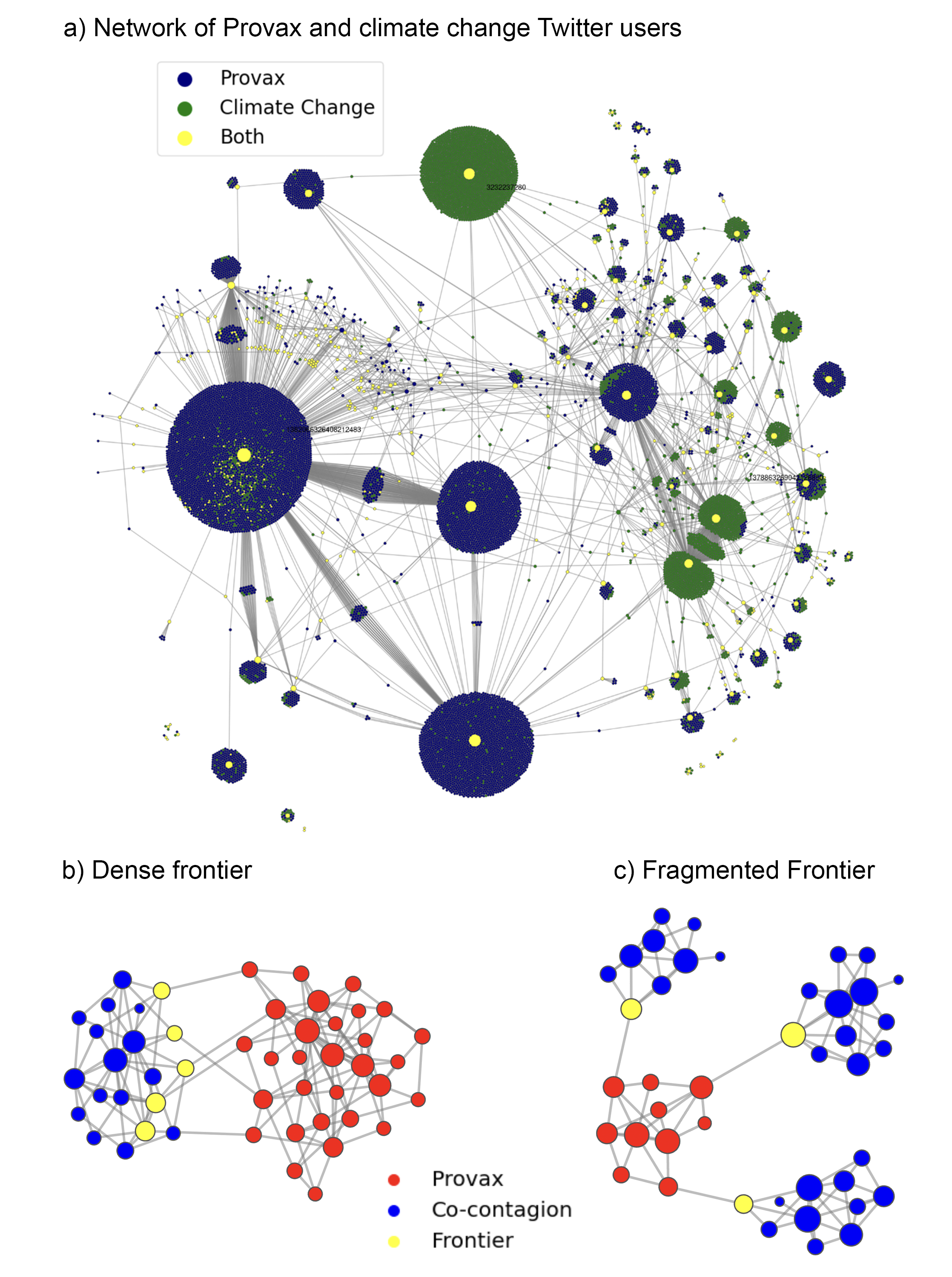}
    \caption{Identifying niche communities for targeted network interventions. Network overlap between users who are pro-vax (blue) and engage with climate change (green) discourse. Those who post about both are denoted in yellow. Different frontiers of diffusion. In b) a dense frontier is observed between pro-vaccination communities, whereas in c) there is a fragmented frontier.}
    \label{fig:vax-climate}
\end{figure}

As an example to introduce the overlapping networks between two contagions, Figure~\ref{fig:vax-climate} visualizes the shared network structure between pro-vax individuals (blue) and those who post about climate change (green) which is one has one of the greatest overlaps with pro-vax users based on switcher timelines. A full list is included in Appendix~\ref{secA2}. Those who post about both are denoted in yellow. One immediate observation is that those who are both pro-vax and discuss climate change (yellow) are local opinion leaders, occupying the central nodes amongst a cluster of nodes. There are a few ways to characterize the different clusters. For instance, Cluster A is predominantly pro-vax, with scattered climate change and both inside (Figure~\ref{fig:vax-climate}a). Cluster B features heterogeneous audiences (both green and blue). Cluster C feature a homogeneous group who discusses climate change. 

The intervention strategies for pro-vaccine conversion (increases in vaccination confidence) can thus be characterized by a trade-off between low-existing converts for high marginal gain and the probability of conversion. Cluster C may have the highest potential yield but conversion may be more difficult due to the lack of network effects; in contrast, Cluster B will have less yield but a higher conversion probability. 

Intuitively, we can therefore generalize these topologies as having \textbf{dense} frontiers or \textbf{fragmented} frontiers. Figure~\ref{fig:vax-climate}a and b show these cases, respectively, and are subgraphs drawn from existing networks. Dense frontiers involve a large, shared interface between the two communities; in this case, one will be the pro-vax community. The other will have a densely connected community of itself. On the other hand, fragmented frontiers feature the pro-vax community interfacing with fragmented communities---the downstream co-contagions are not connected together. This provides us with a dueling hypothesis. On one hand, diffusion on dense communities may be more rapid; on the other hand, they may also be stymied more quickly due to dormancy. To quantify these phenomenon directly, we utilize the larger 2020 COVID-19 pandemic dataset, and provide data-driven network-based intervention simulations.

\subsection{Simulated network interventions}

Behavioral contagions often travel in tandem with each other. Some contagions are synergistic---one paves the way and reduces the resistance for other contagions. Other contagions may increase this resistance by inoculating subjects. The large-scale 2020 COVID-19 pandemic data uniquely captures hashtags and their co-evolution on dynamic networks. Not only can we identifying and measure the extent behavioral contagions are related to pro-vaccine attitudes, we can also simulate and compare different network  interventions.

We first identify the top hashtags that co-occur with pro-vax tweets. We remove broad, COVID-19 related keywords, then cluster the top keywords into specific categories, given below:

\begin{itemize}
    \item Pro-vax: ``getvaccinated'', ``vaccineswork'', ``thisisourshot'', ``vaccineswork'', ``vaccineequity'', ``peoplesvaccine'', ``vaccinessavelives''
    \item Artificial Intelligence: ``ai'', ``rstats'', ``datascience'', ``machinelearning'', ``bigdata''
    \item Climate Change: ``climatechange'', ``climatecrisis'', ``science''
    \item Events: ``tokyo2020''
    \item Other diseases: ``ebola'', ``hiv'', ``malaria''
\end{itemize}

Each of these keyword sets define a network that overlaps the pro-vax network, leading to multiplex networks. Intuitively, their multiplexity of network connections makes sense. Climate change is related to general science skepticism. The Tokyo Olympics of 2020 were first canceled and then extended to the summer of 2021---given the large population of people entering into Japan, vaccine restrictions and masking protocols were taken very seriously. Other diseases---such as Ebola, HIV, and malaria---naturally are vaccine-related topics. More interestingly, artificial intelligence and data science were also popular co-occuring hashtags. We posit this may be related to trust in science or online ``clout-chasing'' strategies, where people utilize unrelated hashtags for increased attention~\cite{chang2022justiceforgeorgefloyd}. The size of the frontier and network are reported in Table~\ref{tab:frontier-size}.

\begin{table}[!htb]
\begin{tabular}{|l|l|l|}
\hline
                               & \textbf{Frontier Size} & \textbf{Total   Network Size} \\ \hline
\textbf{AI and Data Science} & 5,111                    & 10,005                        \\ \hline
\textbf{Tokyo Olympics}      & 4,866                    & 6,580                         \\ \hline
\textbf{Climate Change}      & 5,214                    & 6,729                         \\ \hline
\textbf{Disease}               & 6,009                    & 8,394                         \\ \hline
\end{tabular}
\caption{Frontier size and total network size of co-contagions.} \label{tab:frontier-size}
\end{table}

In light of this quantitative network analysis, we both conceptualize and operationalize the difference between monochrome and multichrome contagions. \textbf{Monochrome contagion} is characterized by a single-category focus that spreads through a targeted or defined network, akin to how diseases propagate primarily within susceptible populations or through particular transmission networks. In contrast, \textbf{multichrome contagion} pertains to broader, more diverse spreading phenomena, where multiple ideas, narratives, or innovations may intermix and cross over between domains, forming multiplex contagion networks.

\begin{figure}
    \centering
    \includegraphics[width=1.0\linewidth]{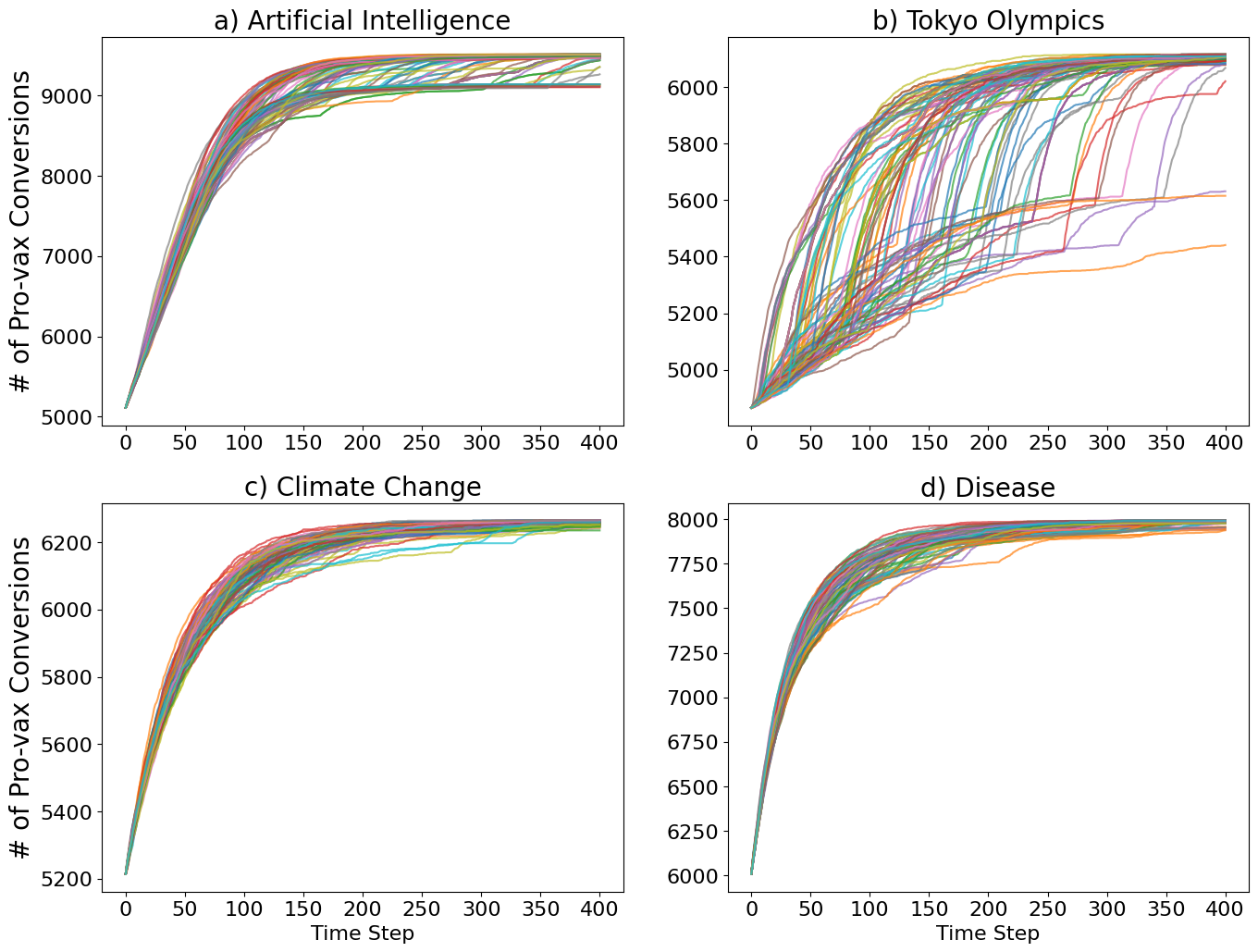}
    \caption{Simulated pro-vaccine diffusion. Panels show individual trajectories on adjacent a) artificial intelligence and data science (n=10,005), b) Tokyo Olympics (n=6,580), c) Climate Change (n=6,729), and d) Disease (n=8,394) networks, over 400 time steps, with a dormancy rate of $\tau = 0$, and a diffusion probability scaling factor of $p = 0.01$.}
    \label{fig:diffusion-no-dormancy}
\end{figure}

Upon defining these multiplex networks, we then conduct network diffusion experiments across a variety of multicrhome contagions (see Algorithm~\ref{diffusAlgo} in the Methods). Figure~\ref{fig:diffusion-no-dormancy} shows the diffusion curves of pro-vax attitudes across four adjacent networks, specifically a) artificial intelligence and data science, b) the Tokyo Olympics, c) Climate Change, and d) Disease. These are done without dormancy ($\tau = 0$) over 400 time steps for illustrative purposes.

For each of these curves, we observe upward motions resembling a Hill function, before saturating at some carrying capacity. However, these transitions are not necessarily smooth. We observe mini "Hill Functions" within each diffusion curve, most evident in Figure~\ref{fig:diffusion-no-dormancy}b, where at certain time steps the number of infected individuals increases significantly. This "step-wise" process of diffusion captures underlying broadcasting phenomenon, where certain users have large audiences and once infected, the become superspreaders of behavior.

\begin{figure}
    \centering
    \includegraphics[width=1.0\linewidth]{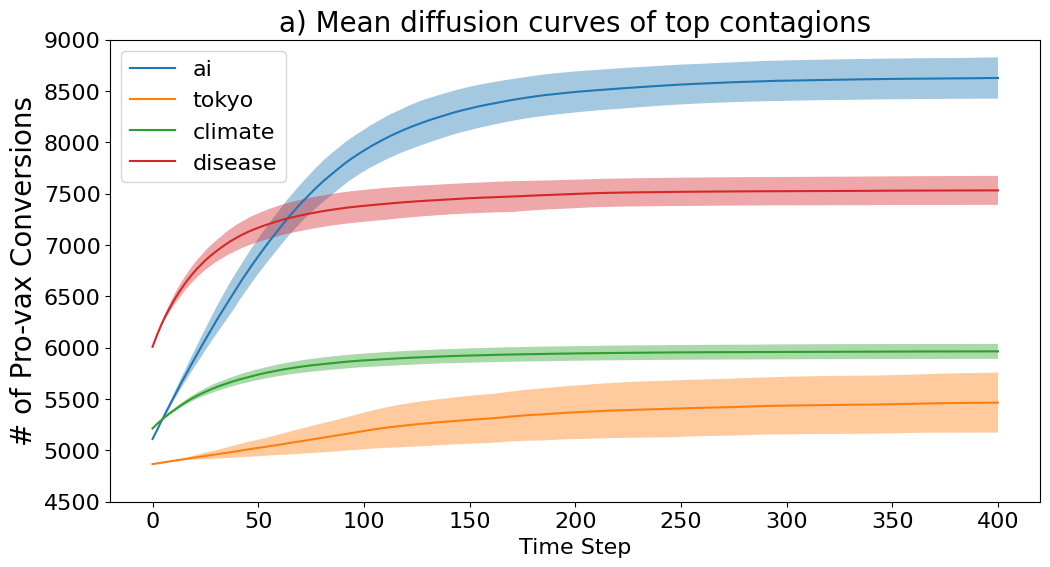}
    \includegraphics[width=1.0\linewidth]{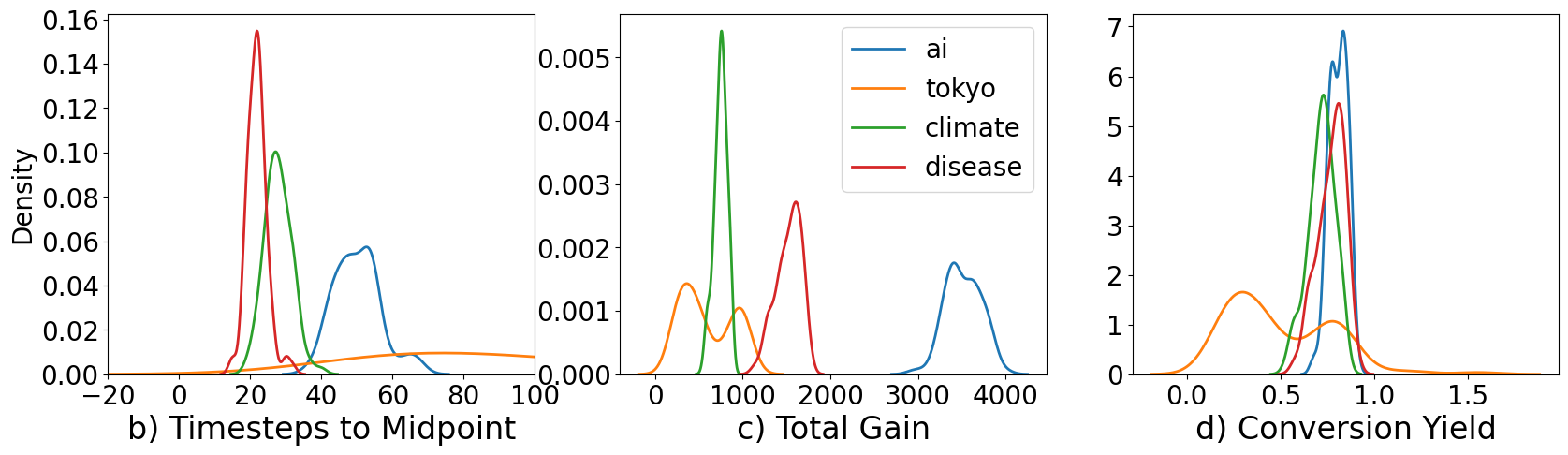}
    \caption{Simulated pro-vaccine diffusion. Shown are trajectories with dormancy on artificial intelligence and data science ($n = 10,005$), Tokyo Olympics ($n = 6,580$), Climate Change ($n = 6,729$), and Disease ($n = 8,394$) networks, over 400 time steps, a dormancy rate of $0.005$, and a diffusion probability scaling factor of $p = 0.01$. Kernel density estimates of a) diffusion speed, b) total diffusion depth, and c) conversion yield for artificial intelligence and data science, the Tokyo Olympics, Climate Change, and disease.}
    \label{fig:diffusion-dormancy}
\end{figure}

In situations where dormancy $\tau$ is zero, the expected outcome is always full diffusion---as time goes by, all users will eventually diffuse the message. Figure~\ref{fig:diffusion-dormancy} shows diffusion with $\tau=0.005$, which means at every time step there is a $0.5\%$ of an infected (viable and active) individual to go dormant. In Figure~\ref{fig:diffusion-dormancy}, AI and data science (blue) produces the highest level of diffusion. However, it's important to note that it does not necessarily diffuse the fastest. Rather, other multichrome contagion with diseases (red) such as Malaria and HIV diffuses faster.

On the other hand, Climate Change (green) features the least amount of variance between diffusion iterations. The network for the Tokyo Olympics (orange) has the lowest diffusion ceiling while also having the greatest variance in diffusion trajectories. To summarize these findings, Figure~\ref{fig:diffusion-dormancy} shows key dependent variables in the diffusion process. Figure~\ref{fig:diffusion-dormancy}a) shows that the monochrome category of disease (red) diffuses the fastest, followed by climate (green), and artificial intelligence (blue). Tokyo Olympics diffuses the slowest with huge variance. 
The total gain also varies, as shown in Figure~\ref{fig:diffusion-dormancy}b), mostly contingent on the total size of the network. The AI community boasts the largest gains, whereas climate change has a consistent level of diffusion. Interestingly, diffusion on the Tokyo Olympics is bimodal, which suggests a bifurcation of results based on whether specific superspreaders go dormant or not.


While we see variance across the total gain, the actual conversion rate---the number converted divided by the total possible---is actually quite consistent across climate change, disease, and AI at around $80\%$. However, the Tokyo Olympics still maintains a bimodal distribution where one mode overlaps with other contagions, and the other mode is below $50\%$. We posit this is related to the nature of the communities behind these multichrome (co-)contagions. The AI and data science, climate change, and disease are all existing communities with significant Twitter presence. On the other hand, the Tokyo Olympics, while garnering significant usage, is a one-time event and temporary in nature.

\begin{figure}
    \centering
    \includegraphics[width=1.0\linewidth]{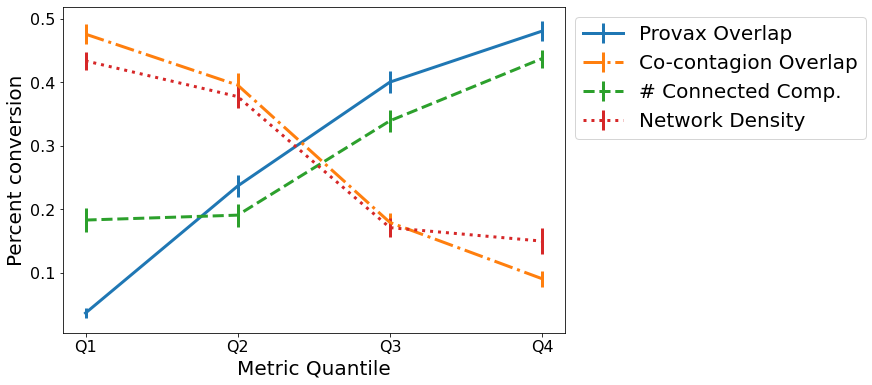}
    \caption{Average pro-vaccine conversion yield based on topic features such as the percentage overlap of the network frontier with the a) pro-vax community and b) secondary hashtag, and also against network features such as c) the number of connected components and d) the network density. Diffusion experiments (100 simulations each) were run over 400 time steps, a dormancy rate of $\tau = 0.005$, and a diffusion probability scaling factor of $p = 0.01$.}
    \label{fig:conversion-yield-stats}
\end{figure}

To validate the robustness of these observed dynamics, we randomly sample $101$ co-occurring hashtags along with the vaccine hashtag, each with $100$ simulations ($n = 101,000$). Figure~\ref{fig:conversion-yield-stats} shows how the conversion yield shifts due to various network statistics. First, we consider the extent of initial overlap. The initial percentage overlap with the pro-vax community (blue), a strong positive correlation is observed---for secondary hashtags that take up a large share over the total of pro-vax users, there is greater diffusion. In other words, communities that significant levels of pre-existing pro-vax users are good targets for intervention. 

However, this is not true for overlap with the secondary community. As the co-contagion community overlaps with the frontier, it lowers the conversion yield. The greater the frontier's overlap is with all the secondary hashtag, the less diffusion occurs. 

It is worth noting that the conversion yield is based solely on viable candidates and is therefore not affected by double counting. Examples of these include monochrome networks of other diseases such as HIV and malaria. Further sufficient differentiation is required. The choice of target intervention community should depend on the expected total yield, rather than simply the potential penetration depth. Therefore, multichrome contagions are more effective at producing desired diffusion of pro-vaccine attitudes.

Outside of topics, we also consider how network topology directly influences diffusion. As these networks may not be fully connected, we rely on network metrics that do not require connected networks as a pre-condition. The green line shows the number of connected components on the percentage conversion, where a greater number of connected components indicates greater fragmentation. Somewhat counterintuitively, as the number of connected components increase, the percentage conversion also increases. This goes against the hypothesis that denser, connected networks can experience greater levels of diffusion. This is likely related to the presence of dormancy; once spreading individuals at the frontier become dormant, the entire network is no longer engaged with pro-vaccine conversations. 

Under this assumption, as expected, network density (red) is negatively correlated with the percentage of conversation (yield), again due to dormancy. We further posit this relates to the broadcasting nature of social networks like Twitter, where superspreaders (opinion leaders) form local star networks with highly tuned audiences. Even in well-connected sub-graphs, these connections are formed between individual broadcasters. Thus, should these broadcasters go dormant, then downstream broadcasters are cut-off from the cascading pro-vaccine messaging. It is therefore better to strategically initiate diffusion with fresh, albeit smaller communities close to the diffusion frontier, rather than with a large, deeply connected one. Both results on network metrics suggest that breadth matters more than depth.

In sum, our results suggest that monochrome contagions are characterized by high co-contagion overlap, high density, and few connected components. In contrast, multichrome contagions have low initial co-contagion overlap, low density, and may contain many connected components. Contagions that intersect in a broad but shallow fashion are the most effective, whereas dense communities may suffer from premature saturation effects if individuals are prone to dormancy. 

\subsection{Wavering individuals are complex}

\begin{figure}
    \centering
    \includegraphics[width=1.0\linewidth]{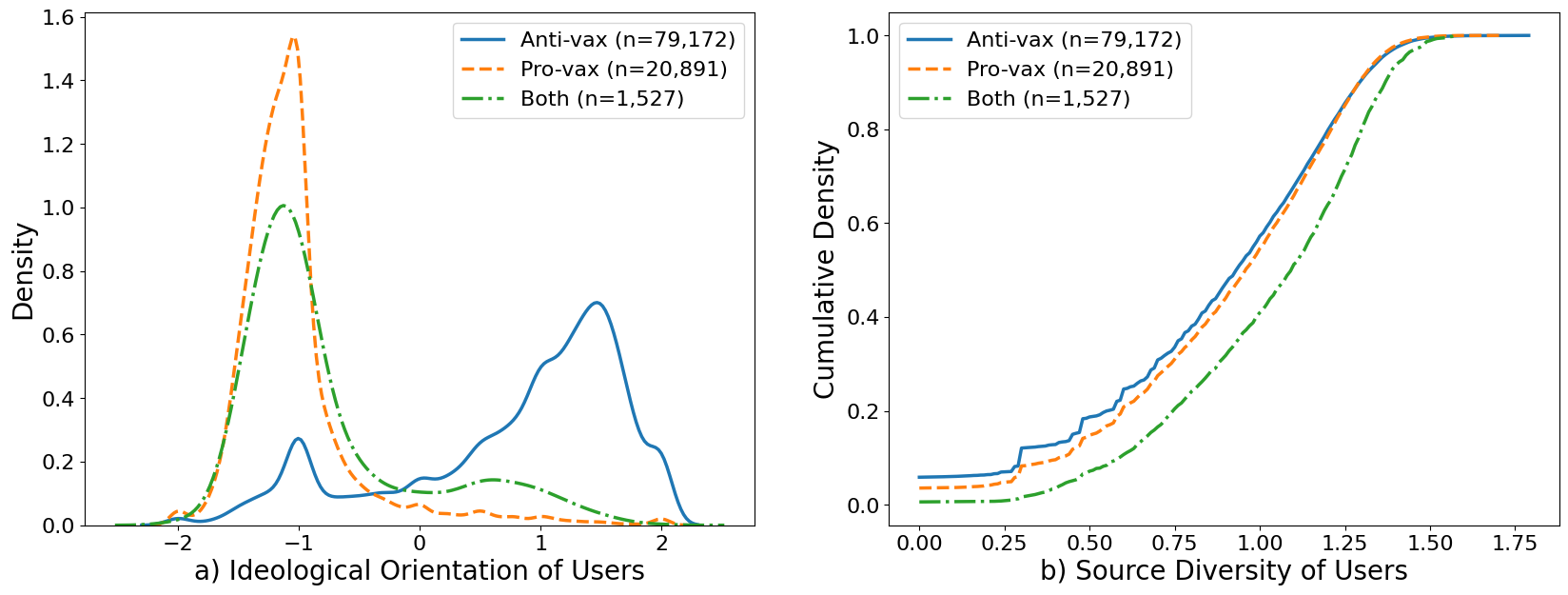}
    \caption{Information profiles of pro-vax (orange), anti-vax (blue), and wavering (green) individuals. a) shows the ideological position of wavering individuals. b) shows the source diversity of their information consumption environments.}
    \label{fig:political-position}
\end{figure}

Having broadly identified the top contagion communities, we conduct a deep dive into those with pro-vaccine, anti-vaccine, and wavering individuals, which, as denoted in the methods, are sampled indivudlas from the greater COVID-19 dataset with their entire timelines extracted. Figure~\ref{fig:political-position}a shows the distribution of political position of pro-vaccine, anti-vaccine, and wavering individuals. The X-axis shows the ideological orientation ranging from $-2$ to $2$. 

Based on the distributions, anti-vax individuals (blue) are predominantly right-leaning, although a small mode centered around -1 indicates a subpopulation is strongly left-leaning as well. Pro-vax (orange) individuals are almost exclusively left-leaning, with one mode centered around 1.2 and a positive tail.

More interestingly, wavering individuals (green) are predominantly left-leaning. However, they also have a heavier tail than pro-vax individuals. This indicates wavering individuals are more ideological heterogeneity on average in comparison to both left- and right-leaning individuals. 

Information can also come from different sources, such as news media or other websites. For every user, we attach a score for source diversity based on Shannon entropy:
\begin{equation} \label{eq:entropy}
    H(x) = - \sum p(x) \log p(x)
\end{equation}
Here, $p(x)$ denotes the probability of an event, and in our case it means the frequency of posting a certain online domain. Thus, a higher Shannon entropy in this context means greater diversity of information diet (consumption). Figure~\ref{fig:political-position}b shows the cumulative distribution; the more rightward (positive) the S-shaped curve is, the more aggregate diversity. Anti-vax individuals (blue) have the least information diversity, but they are comparably close with pro-vax individuals (orange). On comparison, wavering individuals have a much greater skewness by quantile (green). In other words, the average wavering individual will consume, engage, and share from more diverse sources than their pro-vax or anti-vax counterparts, which supports the hypothesis that they are more deliberate decision makers through constructing more diverse information environments. Figure~\ref{fig:network-ideas} in the Supplementary Information shows the top $100$ keywords appearing in the timelines of the wavering individuals.

Beyond the ideology and source, we are also interested in comparing the relative ``importance'' across anti-vax, pro-vax, and wavering individuals. To do so, we leverage social network analysis applied to the retweet network. Formally, a network $G$ is comprised of nodes $V$ and edges $E$. Edges demarcate connections between nodes; they are directed (node $a$ is connected to node $b$ but possibly not vice versa). Here, nodes are individuals and edges are directed edges defined by retweeting from node $a$ to node $b$.

\begin{figure}
    \centering
    \includegraphics[width=0.85\linewidth]{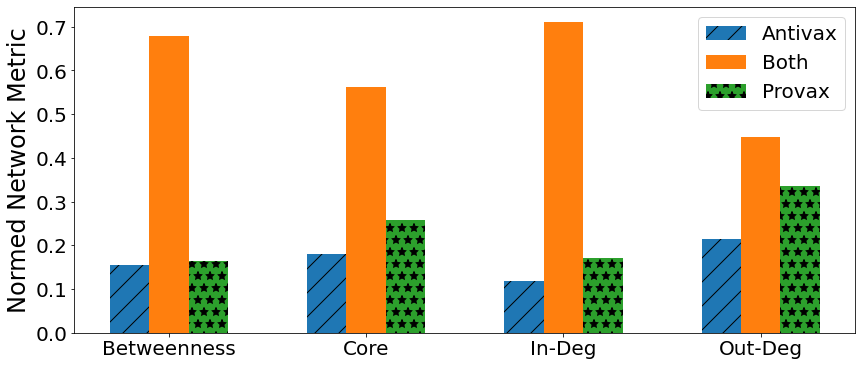}
    \caption{Network measures of pro-vax, anti-vax, and wavering individuals (sample network size $101,590$), including betweenness centrality, core count, in-degree, and out-degree.}
    \label{fig:centrality-measures}
\end{figure}
Figure~\ref{fig:centrality-measures} shows key centrality measures of the retweet network, including betweenness centrality, core count, in-degree, and out-degree. Wavering individuals in all cases measure higher than their counterparts on average. High betweenness indicates they are more effective information brokers and bridges; higher core count indicate they are more central and influential; higher in-degree and out-degree indicate they retweet more and are retweeted more overall. Interestingly, pro-vax individuals have substantively higher out-degree compared to anti-vax individuals, which indicates they are more likely to be retweeted. 

Together, these results demonstrate that wavering individuals are more moderate though left-leaning, engage with more diverse sources and topics, and are more central, influential, and bridging. This evidence paints a picture of more deliberate decision makers who weigh various forms of evidence and engage directly with their information environments. 

\section{Discussion and Conclusion}\label{sec:discussion}

Vaccine hesitancy is a public health challenge driven by factors ranging from lack of trust in healthcare systems to politicization of scientific knowledge~\cite{macdonald2015vaccine,larson2018state,razai2021covid,brewer2017meta}. While recent studies have often framed the COVID-19 context in terms of political partisanship~\cite{chen2021covid}, our findings underscore the need to move beyond a binary pro-vax versus anti-vax lens. Instead, we show that individuals who ``switch'' between acceptance and hesitancy---often termed wavering or hesitant---possess distinct characteristics that position them as crucial influencers within social networks. Our work thereby integrates and extends previous findings on social, cognitive, and structural dimensions of vaccine hesitancy, in particular leveraging social networks and peer influence to drive pro-vaccine interventions~\cite{schmelz2021overcoming}.

How these social networks, particularly on social media, amplify or attenuate vaccine hesitancy has been examined extensively~\cite{puri2020role,lazarus2021global,loomba2021measuring}. Of particular concern is how misinformation and anti-vaccination messages seem to diffuses faster and yields more engagement than pro-vaccine messages~\cite{allen2024quantifying,vosoughi2018spread}. One reason could be the communities between anti-vax and their adjacent groups---such as conspiracy groups~\cite{rathje2022social,pertwee2022epidemic}---have more organic overlaps or targeted campaigns. Our analysis also reveals an under-explored opportunity: switchers frequently engage with multiple issue-areas---e.g., climate change, data science, HIV prevention---creating ``crossroads'', where legitimate public health messages could gain traction if effectively targeted through such hitchhiking effects. These pro-vaccine adjacent groups can potentially drive equal or even greater levels of synergistic diffusion as a result. We thereby extend the literature on how socio-political environments and moral language may shape public health attitudes~\cite{centola2018behavior,amin2017association}. 

Through this understanding of how public health messages spread, especially pro-vaccination sentiments, our study distinguishes between what we term monochrome and multichrome contagions. Monochrome contagions, such as those focusing solely on COVID-19 vaccine information or closely related diseases, enjoy rapid initial diffusion within homogeneous networks already attuned to health issues. Yet, they often saturate quickly due to highly overlapping audiences. This observation builds on the notion that dense networks can facilitate fast but also limit the spread of behavior~\cite{centola2018behavior}. Conversely, multichrome contagions---messages that blend multiple themes or issues (e.g., combining vaccine advocacy with climate change or seemingly unrelated topics such as data science)---demonstrate an unexpected advantage. Although their initial uptake may be slower, these messages diffuse more broadly and sustainably, an outcome that echoes the potential for network-based interventions in heterogeneous communities~\cite{jarrett2015strategies}.

A particularly related and salient insight is the so-called ``niche connectivity paradox'', where fragmented networks sometimes yield higher overall diffusion than more cohesive ones. This appears to happen because smaller, loosely connected clusters remain ``fresh'' audiences less susceptible to dormancy or echo-chamber effects. These findings align with prior work showing that peer influence and social norms are most impactful when they reach untapped segments of the population~\cite{moehring2023providing}. By contrast, tightly knit communities can adopt new behaviors more swiftly but also become saturated---and thus go dormant---much sooner. This phenomenon indicates that while dense clusters excel in short-term campaigns, fragmented networks may sustain longer diffusion cycles, which complement earlier observations about the diversity of influences driving vaccine attitudes~\cite{brunson2013impact,brewer2017meta}. Methodologically, our work is able to circumvent limitations in purely empirical analyses such as the lack of observable counterfactuals~\cite{bloom2014addressing}, while adding crucial  data-driven empirical validity to purely theoretical results~\cite{chang2018co,chang2019co,chen2019imperfect}.

Lastly, by analyzing information diets and social network positions, we demonstrate that wavering individuals are not necessarily poorly informed or politically apathetic. Rather, they exhibit deliberation---a tendency to seek out multiple information sources and engage with diverse topics. Prior research has noted that vaccine hesitancy can emerge from beliefs, behaviors, and social norms~\cite{lewandowsky2013role,toriumi2024anti}, rather than simple ideological predisposition. Our data reveal a similar phenomenon: switchers lean left overall, but maintain broader ideological variance, which indicates vaccine attitudes can transcend rigid partisan lines. Their network centrality, measured by betweenness and core connectivity, further indicates that they may serve as ``information bridges'' capable of spreading or reinforcing messages---both pro- and anti-vaccine---across otherwise siloed communities.

In sum, our data-driven simulation results point to three potential public health interventions in the field. First, we can \textbf{conceive multichrome strategies}. Consistent with literature suggesting integrated health messages are more effective when they resonate with broader values~\cite{centola2018behavior}, we recommend merging vaccine advocacy with other prominent social or scientific issues. Such ``co-contagion'' strategies may minimize resistance by linking vaccination to topics (e.g., climate change) that switchers already prioritize on. Second, we can \textbf{leverage fragmentation of underlying niche connectivity}. Building on the paradoxical finding that fragmentation can aid diffusion, campaigns will likely benefit from identifying loosely connected clusters (so-called micro-influencers), rather than overly dense communities for ``starter'' interventions. This approach can minimize premature dormancy and maximize eventual reach and adoption. Lastly, we can strategically \textbf{target deliberative switchers}. Our findings reinforce existing calls to recognize vaccine hesitancy as an individualized rather than purely ideological phenomenon~\cite{razai2021covid}. By focusing on switchers---who are more deliberate, better networked, and open to multiple sources--- interventions can simultaneously leverage their centrality and responsiveness. 

Our research has some limitations. Our study relies on Twitter data, which, while extensive, may not fully capture offline dynamics or the role of non-digital social ties in shaping attitudes~\cite{jarrett2015strategies}. Future investigation could triangulate multi-platform or offline data to validate the broader relevance of our findings. Additionally, more granular longitudinal analyses could reveal precisely when and why switchers transition between acceptance and hesitancy, aligning with calls for time-sensitive modeling of vaccine decision processes~\cite{brewer2017meta}. Lastly, although our simulation framework points to strategic ways of targeting network communities, field experiments in real-world settings---such as local health campaigns or community forums---are needed to test whether multichrome interventions can indeed drive the desired shift toward reducing hesitancy while increasing vaccination confidence on a large scale.

Characterizations of vaccine attitudes beyond a simple dichotomy are necessary, and the concept of multichrome contagions proposed in this work may pave the way for future effective public health interventions. By recognizing switchers' unique profile---well-connected, topic-diverse, and moderately ideological---policymakers, healthcare providers, and advocacy groups will be able to design more finely tuned network-based strategies that harness the structural and behavioral dynamics of social networks.

\section{Methods} \label{sec:methods}

\subsection{Data collection}

As mentioned in the introduction, we collected, curated and utilized two extensive datasets. The first dataset (Dataset 1) is the largest public Twitter dataset on the COVID-19 pandemic~\cite{chen2020tracking}. Data collection began on January 28, 2020, using Twitter’s streaming API and Tweepy to track trending keywords and accounts from the onset of the pandemic, and produced a data set of 3.5 billion tweets. The second dataset (Dataset 2) is derived from Muric et al. (2021), which consists of a large, public dataset on anti-vaxers and pro-vaxers~\cite{muric2021covid}. Collection began October 18, 2020, using the Twitter streaming application programming interface (API) to follow specific anti-vaccine-related keywords. The dataset includes 250 million tweets dating back to 2009, from 107,000 users, allowing us to identify 5,326,468 inter-user connections.

\subsection{Labeling for ideology}
To label for ideology, we construct a weighted score based on their political diet. When users choose to share information on Twitter, they often include URLs to particular domains. These domains often have a political slant or media-bias, commonly rated on a scale of left, center-left, center, center-right and right. We create an aggregate score for this bias using the dot product of the proportions of each subset and their weights, as shown below:

\begin{equation}
    I(i) = \begin{bmatrix}
           \text{\% left} \\
           \text{\% left-center}  \\
           \text{\% center} \\
           \text{\% right-center} \\
           \text{\% right}
         \end{bmatrix}    \cdot 
         \begin{bmatrix}
           -2 \\
           -1  \\
           0 \\
           1 \\
           2
         \end{bmatrix}   
\end{equation}
This provides a weighted score with 0 in the center, -2 as extremely left, and 2 as extremely right. This method has been used frequently in measuring the ideological slant or political information diets of online users. These definitions were taken from \textit{AllSides} and validated with Media-bias/Fact-check. Both resources assign labels to media
sources based on their perceived position within the US media ecosystem, and have been widely used~\cite{chang2023liberals,huszar2022algorithmic,ferrara2020characterizing,chang2022comparative}. As we have complete timelines from Dataset 1, we label users on their ideologies in this fashion.

\subsection{Pruning out bots}
There is some possibility that users who utilize hashtags are simply bots that tweet a lot. To remove these cases, we utilize Botomometer. Bot detection is a fundamental method in assessing social media manipulation. We utilize the latest version of Botometer (version 4). Botometer is a machine learning-based application programming interface produced and maintained by Indiana University~\cite{yang2022botometer}. Both the prior version and current version use an ensemble classifier (a probabilistic mixture of different machine learning classifiers) to output a botscore, which indicates the likelihood a user is a bot or a human. For individual user, Botometer extracts over 1,000 features per user using their most recent 200 tweets and the user's profile to make an inference. This includes an account's profile, friends, temporal activity, language, and sentiment. The models are trained then on human annotated users that indicate the likelihood a user is a social bot. 

For this paper, we use the raw scores, which indicate exactly this likelihood. The distribution of Botscores within our dataset is visualized in Fig.~\ref{fig:bots}, which shows a natural bimodal break between 0.5. Hence, we label users with botscores above 0.5 as bots. Details can be found as shown in Fig.~\ref{fig:bots} in Appendix~\ref{secA1}.

\subsection{Network diffusion simulations for monochrome/multichrome contagions}

Our network diffusion experiment can be summarized in the following algorithm.  Let $ADJ$ be the adjacency matrix of the network defined by contagion $A$ and contagion $B$. Suppose there are $N$ total nodes. Let $Inf$ represent a vector of infected (meaning the number of converted pro-vaccine individuals). Let $\tau$ be the dormancy rate (meaning individuals losing interests in discourse engagement of the topics) and $p$ be a scalar for the diffusion rate. The diffusion experiment can then be effectively simulated as follows:

\begin{algorithm}
    \caption{Network diffusion algorithm.}
    \label{diffusAlgo}
    \begin{algorithmic}
        \State infected[t] $\gets$ sum( Inf )
        \State dorm $\gets \vec{0}$
        \For{$t \in T$} 
            \State dorm += Uniform(0,1) $< \tau$
            \State Viable = $Inf$ $\cdot$ (1-dorm)
            \State $deg_{inf}$ = (ADJ $\times $ Viable).sum(axis=1)
            \State prop = ( $deg_{inf}$ / $deg$ ) * (1 - inf)
            \State $\Delta_{inf}$ = Uniform(0,1) $<$ prop $\cdot p$
            \State Inf += $\Delta_{inf}$
            \State infected[t] $\gets$ sum( Inf )
        \EndFor
    \end{algorithmic}
\end{algorithm}

Over $T$ time steps, the simulation tracks the total number of infected (pro-vaccine conversions), viable, and dormant individuals. At each time step, the number viable users suitable for contagion/transmission is calculated as the infected users minus those who have gone dormant. The number of viable users is then divided against each user's degree, with the diffusion probability proportional to this value~\cite{chang2018co,valente1996social,chang2019co}. 

\backmatter



\bmhead{Acknowledgements}
Supported in part by a research grant from Investigator-Initiated Studies Program of Merck Sharp \& Dohme Corp. The opinions expressed in this paper are those of the authors and do not necessarily represent those of Merck Sharp \& Dohme Corp.


%
%
%
 
\newpage

\begin{appendices}

\section{Bot scores for data processing}\label{secA1}

\begin{figure}[!htb]
    \centering
    \includegraphics[width=0.5\linewidth]{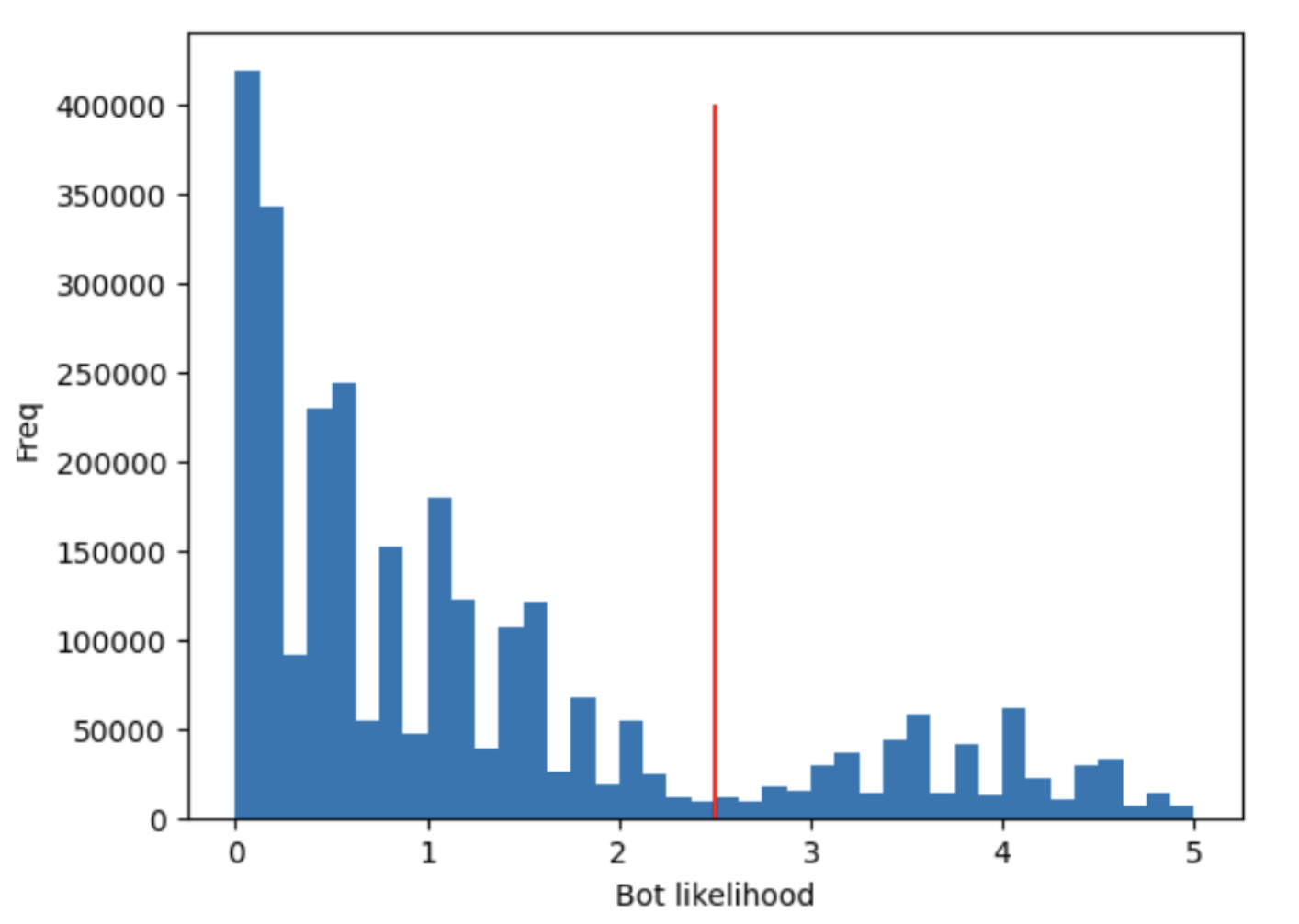}
    \caption{Bot distribution of Dataset 1. Botscores are bimodal which provide a natural break for bot classification.}
    \label{fig:bots}
\end{figure}

\section{Issue Space and Top hashtags}\label{secA2}

Top hashtags can be divided into these top three topics by ideology in Table~\ref{tab:top-topics}. For left-leaning individuals, this includes topics about nature, religion, and choice (this is shared on the right as well). For right-leaning individuals this consists of partisan messaging, distrust in institutions, and ones decrying safety concerns related to the vaccines.
\begin{table}[!htb]
\begin{tabular}{|p{6cm}|p{6cm}|}
\hline
\textbf{Left-leaning}                                               & \textbf{Right-leaning}                               \\ \hline
Nature-oriented individuals (\#ClimateChange, \#Nature, \#WildLife) & Political (\#Trump, \#maga)                          \\ \hline
Religious Communities (\#noisundays)                                & Distrust in Institutions (\#PharmaFraud, \#CDCFraud) \\ \hline
Choice, consent, and rights (\#InformedConsent, \#MyBodyMyChoice)   & Safety Concerns (\#VaccinesKill)                     \\ \hline
\end{tabular}
\caption{Top hashtag categories from individual timelines.} \label{tab:top-topics}
\end{table}

\begin{figure}[!htb]
    \centering
    \includegraphics[width=0.5\linewidth]{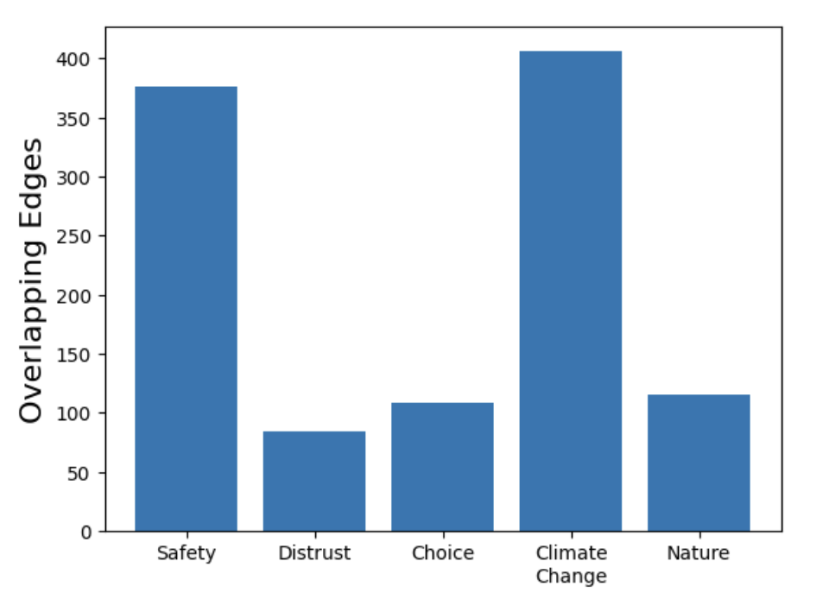}
    \caption{Number of overlapping edges with pro-vax messages.}
    \label{fig:provax-overlap}
\end{figure}

\begin{figure}
    \centering
    \includegraphics[width=0.7\linewidth]{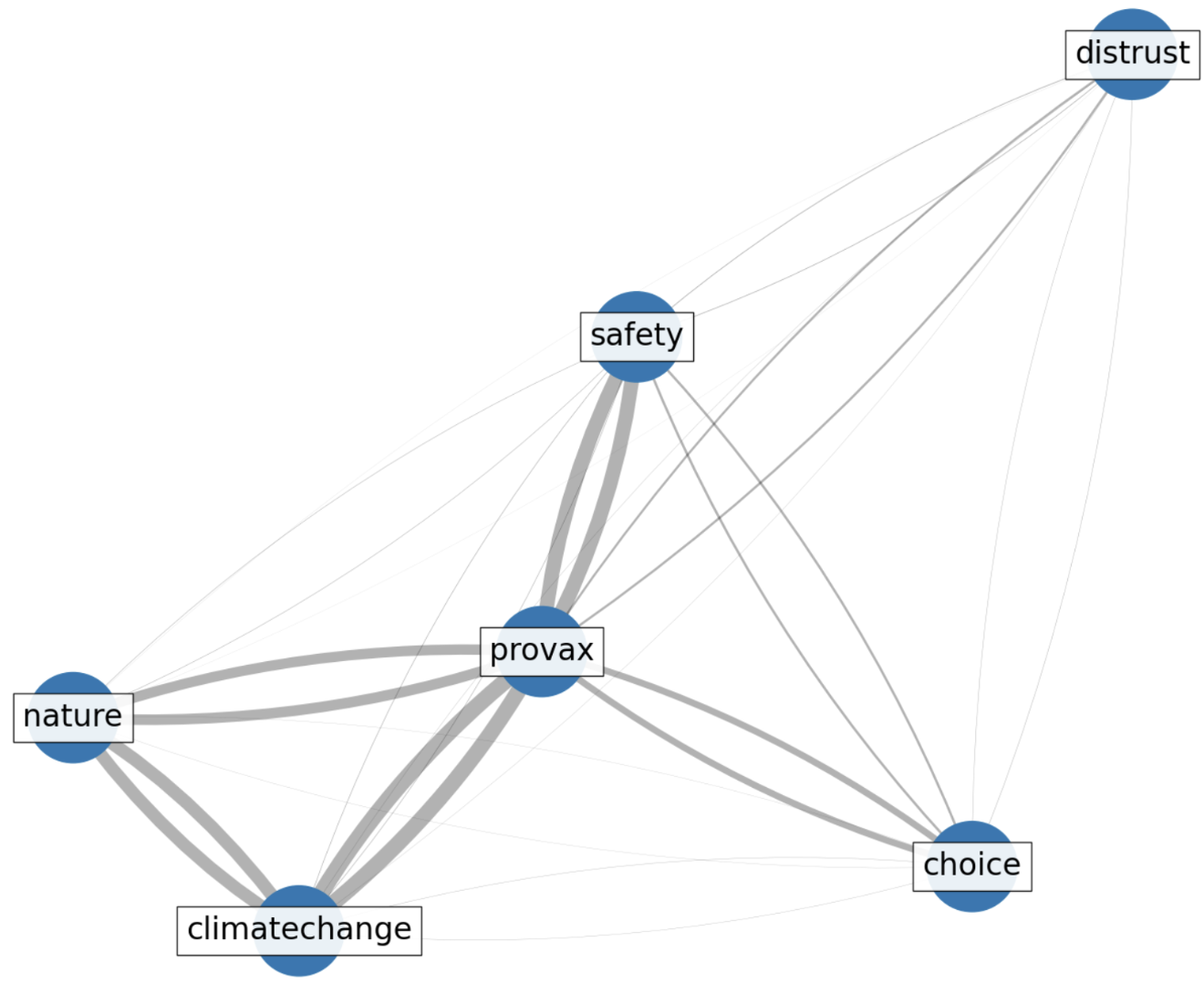}
    \caption{State-machine of transitions between discussion topics.}
    \label{fig:state-machine}
\end{figure}

In particular, we classify vaccine hesitancy into the three broad categories of 1) choice, 2) distrust in institutions and c) concerns about safety. Figure~\ref{fig:state-machine} provides a more granular view of the potential transitions between topics. Unlike simple co-occurrence, we also consider the temporality using state-machines. State-machines are core concepts that describe how actions stimulate transitions into various states.  

Pro-vax discourse occupies the center node, with frequent transitions to climate change and nature. Of the three categories for anti-vax narratives (safety, choice, and distrust in institutions), it also features the most transitions with safety concerns and moderate transitions with choice-based arguments. This indicates individuals that express safety concerns may be more amendable to conversion, and to a lesser extent choice-based argumentation. In contrast, those who express distrust in institutions are the least amendable to conversion.

\begin{figure}
    \centering
    \includegraphics[width=0.6\linewidth]{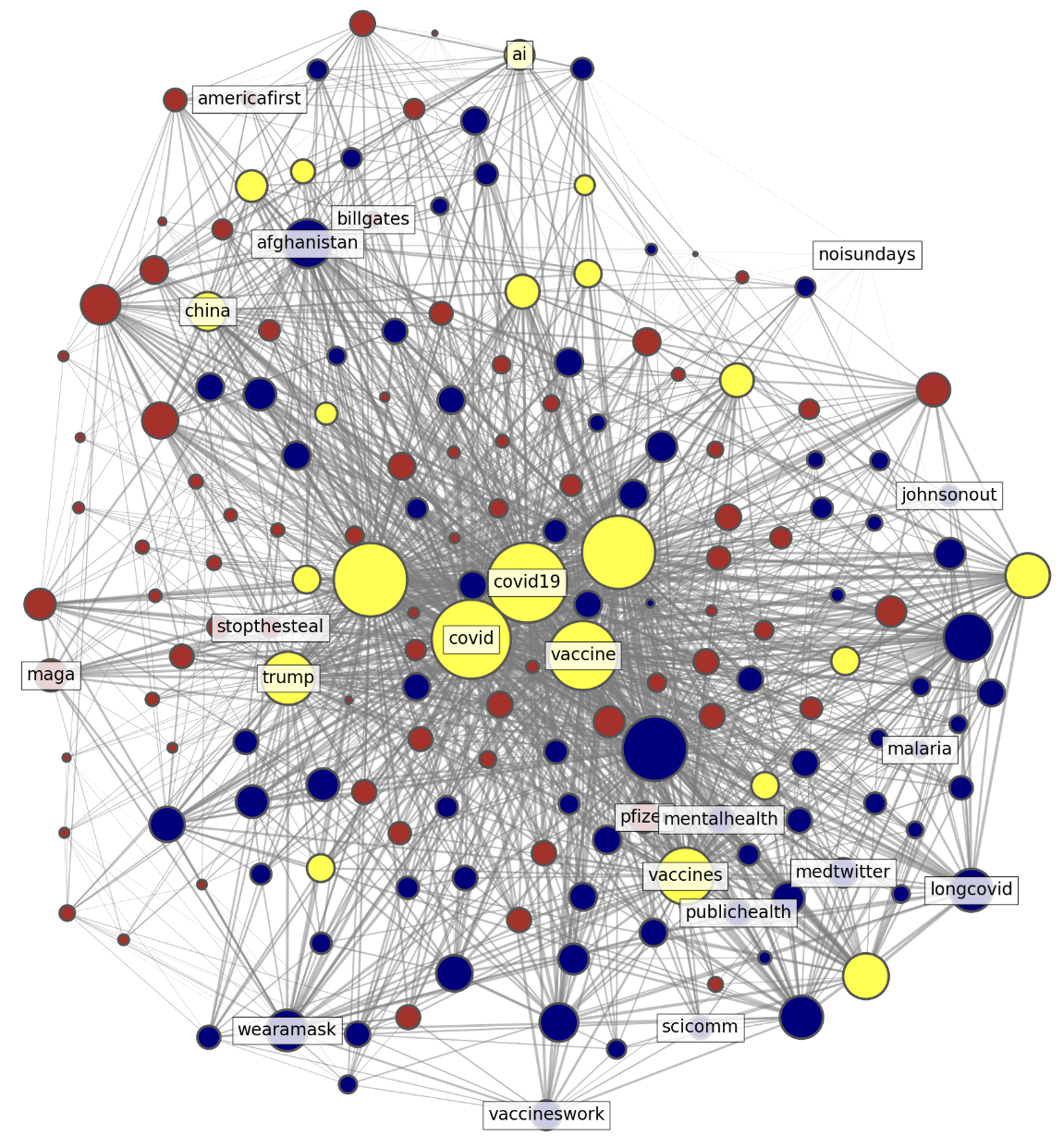}
    \caption{Idea space of wavering individuals is heterogenous}
    \label{fig:network-ideas}
\end{figure}

Red nodes indicate top topics used by anti-vax individuals; blue nodes indicate top topics used by pro-vax individuals. Yellow nodes are neutral, shared words such as ``COVID-19'' or ``vaccination.'' The nodes are sized by the frequency of occurrence. As one might expect, the shared words are the most central (i.e., COVID-19) and used most frequently. However, there is no clear clustering of blue and red words, which indicate mixing within the timelines of wavering individuals. The semantic space for which wavering individuals engage with are also more heterogenous compared to pro-vax and anti-vax individuals, which would be blue-yellow and red-yellow respectively.

\section{Model of synergistic intervention for multichrome contagions}\label{secA3}

The goal of our simulations is to demonstrate potential synergy multichrome contagion. In this context, the diffusion of contagion $A$ (i.e., pro-vaccine sentiment) can be boosted, given its synergy with the co-spreading contagion $B$ (i.e., climate change). 
In our model, a broadcaster $i$ are weakly linked in network $G$. Each broadcaster has an audience (followers), which is a fully connected network. As such, we can define each audience solely by its size $n_i$. The number of existing converts of $A$ is defined as $m_i$. For every contagion/conversion (i.e., pro-vaccine messaging), its diffusion is defined by transmission rate $p$ and dormancy rate of $\tau$. In the empirical and theoretical literature, the dormancy parameter models the eventual saturation of contagion behavior.

We first analyze the diffusion rate within each audience. At the first time step, there are $m_i p$ new converts, and $m_i \tau$ individuals that grow dormant. Thus, the active individuals is $m_i(1+p-\tau)$. As such, the number of viable diffusers $v$ at time $t$:
\begin{equation}
    v_t = m_i (1 + p - \tau) 
\end{equation}
We can then use geometric series to compute the expected cumulative diffusion assuming a fully-connected graph. From geometric series, we know that if $| p - \tau | \geq 1$ then the sum does not converge. Thus, for all cases $p \geq \tau$, we will achieve full diffusion ultimately. Thus, we only consider when $p<\tau$, where the likelihood of dormancy dominates the likelihood of diffusion.

\begin{equation}
    \begin{aligned}
        V &= p m_i + p m_i(1+ p-\tau) + p m_i(1+p-\tau)^2 + \cdots \\\
        &= p m_i \sum_{j=0}^{\infty} (1+p-\tau)^j \\\
        &= \frac{p m_i}{\tau - p}
    \end{aligned}
\end{equation}

This represents the maximum capacity derived from a $m_i$ starting point. Thus, for any audience/follower size of $i$ (denoted by $V_i$ hereafter), we have:
\begin{equation}
    V_i = \min \bigg(n_i, \frac{p m_i}{\tau - p} \bigg)
\end{equation}

Now, we consider two central questions. First, \textbf{who} do we choose to maximize diffusion? Second, what \textbf{topic} should we choose to maximize diffusion? We begin with the first one. If the broadcasters feature no connections, then the best strategy is to choose the maximum carrying capacity:
$$
\max \{ V_i \}_{i=0}^k
$$
A second possibility is all broadcasters are connected. The expected diffusion $y(i)$ starting with the local opinion leader $i$ can be then expressed as:
\begin{equation}
    y(i) = V_i + \sum_{j \neq i}^k p V_j
\end{equation}
where each component is the probability of diffusing into a neighboring broadcaster multiplied by their carrying capacity. However, in this case, our optimal strategy is still to choose community $i$ with the greatest $V_i$, as it strictly dominates all other terms.

The second related question is how to choose the topic. This choice depends on three features, as demonstrated in our main text: the diffusion rate, the network, and maximum carrying capacity (niche/community size). While the network certainly depends on the underlying topic, these network shapes are often very similar in the wild. More importantly, the diffusion rate $p$ and empirical carrying capacity are in direct trade-offs. The greater the overlap, the more synergistic the diffusion but the lower the capacity; the smaller the overlap, the lower the diffusion synergy but the greater the potential for conversion (percentage-wise).


\end{appendices}

\bibliography{sn-bibliography}

\end{document}